\documentclass[journal]{IEEEtran}
\usepackage{graphicx, subfigure}
\usepackage{amsmath}
\usepackage[colorlinks,linkcolor=blue]{hyperref}
\usepackage{cite}

\ifCLASSINFOpdf
\else
\fi
\begin{document}
%
\title{MDVSC---Wireless Model Division Video Semantic Communication}
%
%
%

\author{Zhicheng~Bao,~Haotai~Liang,~Chen~Dong*,~Cong~Li,~Xiaodong~Xu,~\IEEEmembership{Senior~Member,~IEEE},~Ping~Zhang,~\IEEEmembership{Fellow,~IEEE}
\thanks{Parts of this work have been submitted to IEEE Global Communications Conference 2023. \cite{Bao2023MDVSCW}

This work is supported by the National Key R\&D Program of China under Grant 2022YFB2902102.
}}

%
%

\markboth{Journal of \LaTeX\ Class Files,~Vol.~14, No.~8, August~2015}%
{Shell \MakeLowercase{\textit{et al.}}: Bare Demo of IEEEtran.cls for IEEE Journals}
%



\maketitle

\begin{abstract}
This paper introduces a novel method for transmitting video data over noisy wireless channels with high efficiency and controllability. The method derivates from model division multiple access (MDMA) to extract common semantic features from video frames. It also uses deep joint source-channel coding (JSCC) as the main framework to establish communication links and deal with channel noise. An entropy-based variable length coding scheme is developed to adjust the data amount accurately and explicitly. We name our method as model division video semantic communication (MDVSC). The main steps of our approach are as follows: first, video frames are transformed into a latent space to reduce computational complexity and redistribute data. Then, common features and individual features are extracted, and variable length coding is applied to further eliminate redundant semantic information under the communication bandwidth constraint. We evaluate our method on standard video test sequences and compare it with traditional wireless video coding methods. The results show that MDVSC generally surpasses the conventional methods in terms of quality metrics and has the capability to control code length precisely. Moreover, additional experiments and ablation studies are conducted to demonstrate its potential for various tasks. 
\end{abstract}

\begin{IEEEkeywords}
Semantic communications, video transmission, latent transform, joint source-channel coding, variable length coding, model division multiple access.
\end{IEEEkeywords}

%
\IEEEpeerreviewmaketitle

\section{Introduction}
%
%
%
%

\begin{figure*}
  \centerline{\includegraphics[width=6.8in]{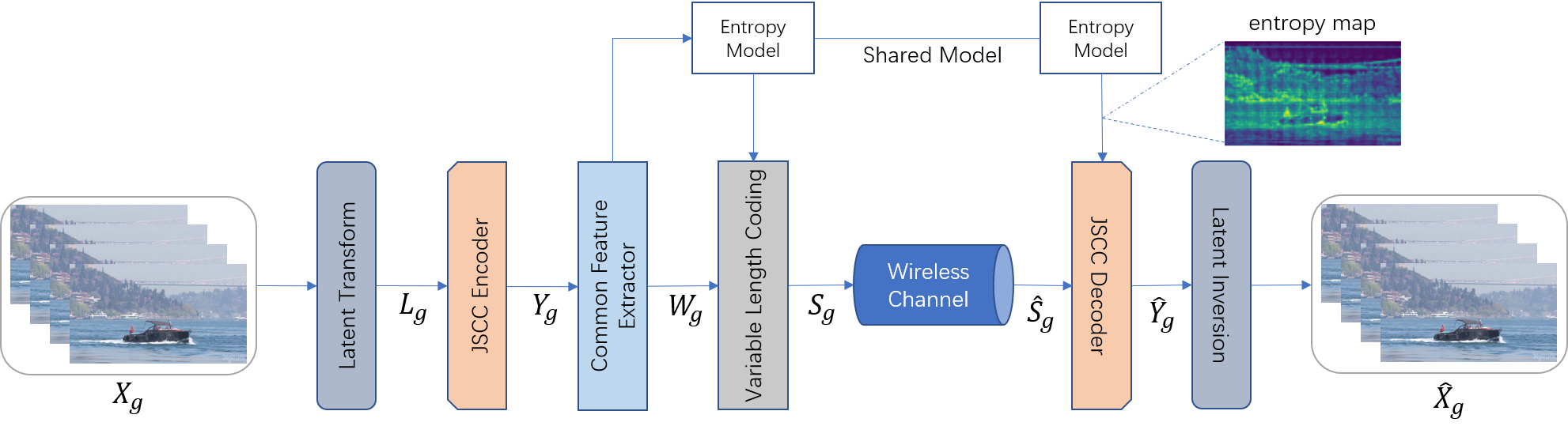}}
  \caption{The framework of the MDVSC. The structure of the transmitter and the receiver is not equivalent, making the receiver take less computing burden. The MDVSC takes an entire GOP as the input. According to the shared entropy model, the receiver is able to reconstruct the video frame from the variable length code words sent from the transmitter.\label{fig1}}
  \end{figure*}
  
\IEEEPARstart{V}{ideo} content is becoming a primary source of internet traffic, thanks to the advances in communication technology and video compression algorithms. However, this also poses a challenge for the communication link that needs to transmit video data efficiently and robustly. Most existing wireless video communication systems use source coding and channel coding separately \cite{Shannon1948AMT}. Source coding, such as H.26X series coding, applies various video compression algorithms to remove redundancies in the video sequences and then converts them into bit sequences through entropy coding. Channel coding, such as low-density parity-check (LDPC) coding, adds parity check bits, which are redundancies, to the bit sequences to cope with distortions caused by the noisy wireless channel. The separated scheme has the advantage of decoupling source and channel coding, making optimizing their performance individually easier.

However, the separated scheme also has some limitations, especially when the demand for higher wireless video transmission quality under lower communication resource costs increases. One of the limitations is the cliff effect caused by the time-varying channel. This not only means that when the channel condition worsens beyond a certain threshold, the video communication receiver cannot distinguish the signal from the noise. Moreover, the entropy coding used by the source coding depends heavily on the accurate estimation of the marginal distribution of the source. Therefore, any error that is not corrected by the channel coding can result in catastrophic error propagation during entropy decoding and severely degrade the performance of the whole video communication system \cite{Rissanen1981UniversalMA, Ball2019IntegerNF}. Another limitation is that the wireless channel bandwidth allocated for one user equipment (UE) in practical video transmission links is fixed during a time and cannot change freely. However, the bit rate of a video can vary greatly and sometimes exceed the bandwidth constraint. If the local video buffer is insufficient, the jitter may occur and affect user experiences.

To optimize the transmission efficiency, joint source-channel coding (JSCC), which combines source processing and channel processing, was studied in \cite{Fresia2010JointSA}. However, traditional JSCC schemes rely on handcrafted arithmetic, and their optimal maximum-likelihood detection is generally an NP-hard problem \cite{Guyader2001JointST}. Moreover, they cannot deal with the cliff effect. With the development of deep neural networks (DNNs), a new JSCC scheme based on DNNs has emerged. Using the ability of complex feature extraction, DNNs have enabled the construction of a communication system. Works in \cite{Bourtsoulatze2019DeepJS, Kurka2019DeepJSCCfDJ, Kurka2020BandwidthAgileIT, Yang2021DeepJS} build a JSCC image transmission system, and its performance exceeds the traditional structure. Based on that, semantic communication is proposed as a new generation of the communication system \cite{Zhang2021TowardWA, Lu2021RethinkingMC, Luo2022SemanticCO}. Works in \cite{Xie2020DeepLE, Xie2020ALD, Dong2023SemanticCS} are the pioneer in implementing the semantic communication system. By using DNNs, source data can be directly encoded as continuous value symbols to be transmitted through a noisy wireless channel, thus overcoming the cliff effect and improving communication system performance. Following that, the authors in \cite{9953110} design a deep joint source-channel coding method to achieve end-to-end video transmission over wireless channels. It is based on the traditional motion estimation and motion compensation (MEMC) framework \cite{Lu2018DVCAE}. Meanwhile, it uses non-linear transform, context-driven semantic feature modeling \cite{Li2021DeepCV}, and rate-adaptive semantic feature transmission to outperform classical H.264/H.265 combined with LDPC and digital modulation schemes. All of these previous works have explored a way to solve the problem of the cliff effect and improve transmission efficiency.

However, there are still some limitations to the video communication task. In general, the current video semantic communication system, such as \cite{9953110}, is based on the traditional video communication architecture, such as the MEMC system. This system captures video frame correlation at the time domain and uses these properties to compress temporal redundancy. One of its operations, optical-flow-estimation in the pixel domain, highly depends on the input dimension. For videos with high resolution and deep color depth, the time complexity for joint coding can be too high to meet the requirement for real-time video transmission. Moreover, the original pixel distribution that humans can perceive intuitively at the pixel domain may not be well captured by the transmitter and represented as transmitting symbols effectively. Furthermore, the delay jitter mentioned above still needs to be considered seriously. Therefore, we are motivated to design a new video communication scheme that can overcome the cliff effect, improve transmission efficiency and have the ability to control data amount precisely.

According to \cite{Zhang2023ModelDM}, semantic features extracted from the same semantic communication model have some similar components, which are called shared information. The remaining components are personalized information. Inspired by this, it is hypothesized that semantic features extracted from the video frames at continuous time domain have more shared information, which is called here as common features. Likewise, personalized information is called individual features. In this case, common features of the whole group of pictures (GOP) and individual features of each video frame are the units to be processed instead of the traditional optical flow and residual information. That is, our system replaces the MEMC system and is based on the common-individual features. Therefore, based on the inspiration mentioned above, the basic framework of the MDVSC is proposed. It is noted that the definition of GOP in our system is not exactly the same as the one in the typical video communication algorithm. It will be explained in more detail in the following section.

Furthermore, to solve the delay jitter mentioned above, one of the solutions is to change the code length according to the channel condition, ensuring that the communication rate is always lower than the channel capacity. However, many variable length coding schemes in semantic communications cannot control code length manually and precisely. In \cite{9953110}, for example, the transmitter implicitly learns and controls the code length. In this case, for a given limited communication condition, such as a fixed channel bandwidth ratio, it is hard to precisely transmit as much data as possible while satisfying capacity. Therefore, in MDVSC, an entropy-based variable length coding scheme is proposed. It introduces entropy as a metric to measure the importance of each symbol. Based on that, code words outputted from the encoder can be further processed according to the demand to explicitly control code length and improve the system's communication flexibility.

To be specific, the contribution of this paper can be summarized as the following:

(1) MDVSC Framework: We propose a new wireless video communication scheme named model division video semantic communication system (MDVSC). It exploits latent space transform, deep joint source-channel coding (JSCC), commonness capture, and variable length coding to establish an efficient end-to-end learnable communication link.

(2) Common and Individual Features Extractor: A simple yet effective network module common feature extractor is embedded in the MDVSC to extract common and individual features. The common and individual features based MDVSC can improve communication performance significantly under severe channel bandwidth limitations.

(3) Entropy-based Variable Length Coding: By introducing entropy as a method of measuring the importance of semantic information, symbols in the MDVSC are not of equal value. Instead, each symbol has its own significance to the performance of video communication. Based on that, variable length coding that can adjust automatically according to the threshold is realized. 

(4) Performance validation: The performance of the proposed MDVSC system is verified across standard video source sequences. As shown in the test results, the MDVSC can achieve better coding gain on quality metrics compared to the traditional wireless video communication system in typical situations. Extended experiments and ablation studies further demonstrate the potential of its particular application. Meanwhile, the MDVSC shows excellent performance in controlling the code length.

The rest of this paper is arranged as follows. In the next section II, the overall system model and the proposed method are introduced. Then, network architecture and corresponding algorithms are detailed in section III. The experiment part will be shown in section IV, and section V concludes this paper. 

 




\section{The proposed method}
In this section, the system model of wireless video communication will be presented first. Following that, the whole framework of MDVSC will be introduced. Then, the mechanism of the model division video transmission will be specified. After that, the entropy-based variable length coding will be shown, and the optimization goal of MDVSC will be derived in the end.

\subsection{System model}

\begin{figure}
  \centerline{\includegraphics[width=3.4in]{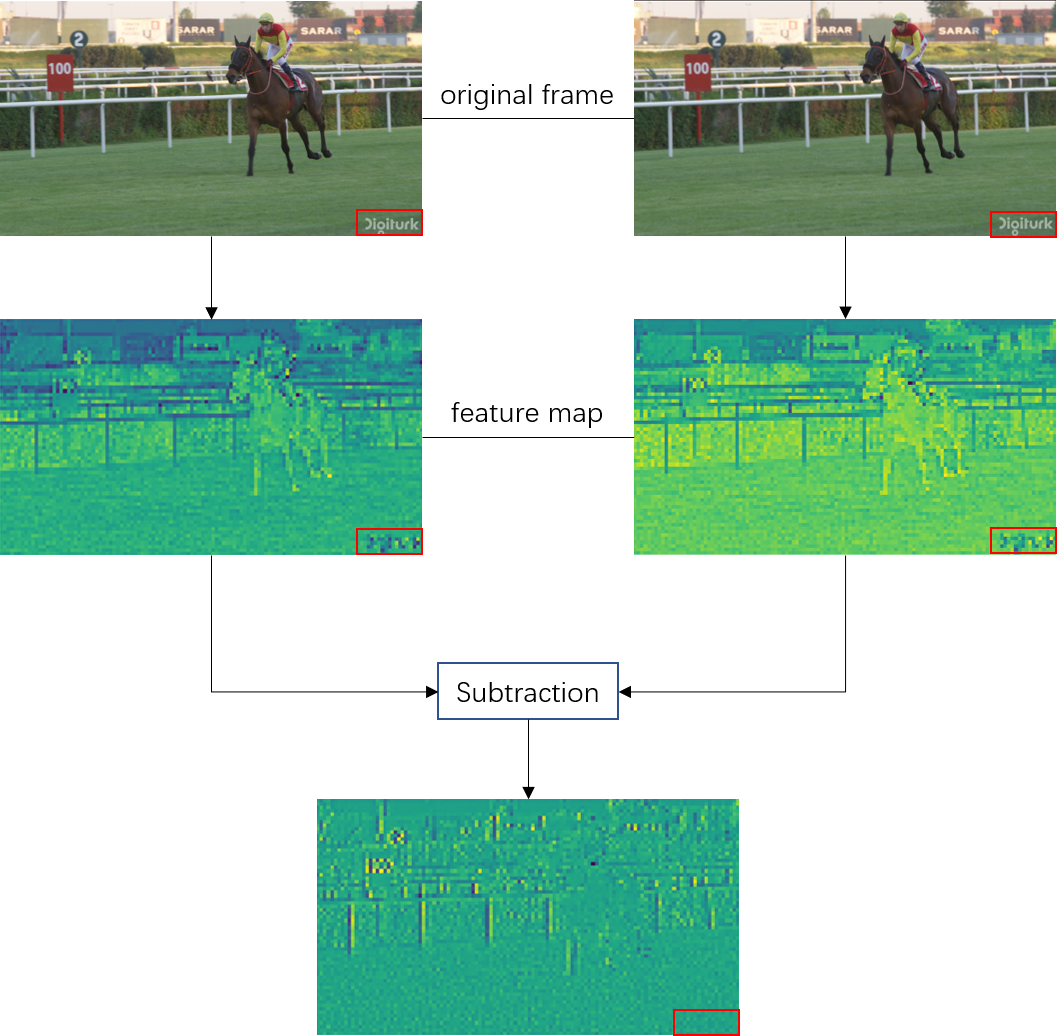}}
  \caption{The visualization of feature maps. After subtraction, common features get removed, and individual features are left. It is clear that most similar features get removed, proving that consecutive video frames have many common features.\label{fig4}}
  \end{figure}
  
Assuming that there is a video sequence to be transmitted
\begin{equation}
    X = \{x_1, x_2, ..., x_T\},
\end{equation}
where the frame at time step $t$ is a vector of pixel intensities in a certain space dimension, namely $x_t \in R^m$. In the semantic communication system, this sequence is divided into some GOPs, i.e., a GOP which contains certain frames can be represented as
\begin{equation}
    X_g = \{x_1, x_2, ..., x_N\},~\text{with}~N\in\mathbf{N}
\end{equation}
 and $x_n$ is the $n$th frames in a single GOP $X_g$. MDVSC takes a GOP like this as the input. After encoding, the GOP $X_g$ is transformed into a sequence of variable-length continuous-valued channel input symbols
 \begin{equation}
      \mathcal{S}_g = \{s_1, s_2, ..., s_N, s_{N+1}\}.
 \end{equation}
For each element $s_n$ in $S_g$, it has
\begin{equation}
\begin{split}
    s_n \in W_{gi}~(1\leq n \leq N)\\
    s_n \in W_{gc}~(n = N+1)\\
    \text{with}~s_n \in R^{k_n}~~~~~&
\end{split}
\end{equation}
where $k_n$ denotes the $k_n$-dimensional channel input. For $1\leq n \leq N$, $s_{n}$ are viewed as the individual features for each frame $x_n$. While $s_{N+1}$ is the common feature symbol $W_{gc}$ of the entire GOP. Details about it will be presented in the next subsection. To measure the cost of communication resources, the channel bandwidth ratio (CBR) is proposed in \cite{Kurka2019DeepJSCCfDJ}. Here it is used for measuring the communication cost of transmitting GOP $X_g$ and is defined as 
\begin{equation}
    CBR = \sum_{n = 1}^{N+1} \frac{k_n}{m}.
\end{equation}
Then the encoded GOP $\mathcal{S}_g$ is transmitted through the noisy wireless channel. The channel is modeled as a transfer function $\mathcal{W}(\cdot~;\mathcal{P})$, and $\mathcal{P}$ denotes the channel parameters. Therefore the received GOP can be formulated as
\begin{equation}
    \hat{S_g} = \mathcal{W}(\mathcal{S}_g;\mathcal{P}) = \{\hat{s}_{1}, \hat{s}_{2}, ..., \hat{s}_{N}, \hat{s}_{N+1}\}.
\end{equation}
Where $\hat{s}_n$ is the common or individual feature affected by channel noise corresponds to $s_n$. For the receiver, it decodes this distorted GOP and tries to recover $\hat{x_n}$ from the noisy $\hat{s_n}$, namely,
\begin{equation}
    \hat{X}_g = \{\hat{x}_1, \hat{x}_2, ..., \hat{x}_N\}.
\end{equation}
So far, the process of a GOP is complete.

It has to be highlighted that in our system, the term group of pictures (GOP) refers only to a stack of video frames. In contrast, some typical video communication algorithms define a GOP as an intra-coded picture (I-frame or keyframe) followed by some predictive coded frames (P-frames). Our communication system does not use the concepts of I-frame or P-frame. Instead, all frames in one GOP are treated equally.

\begin{figure*}
  \centerline{\includegraphics[width=6.8in]{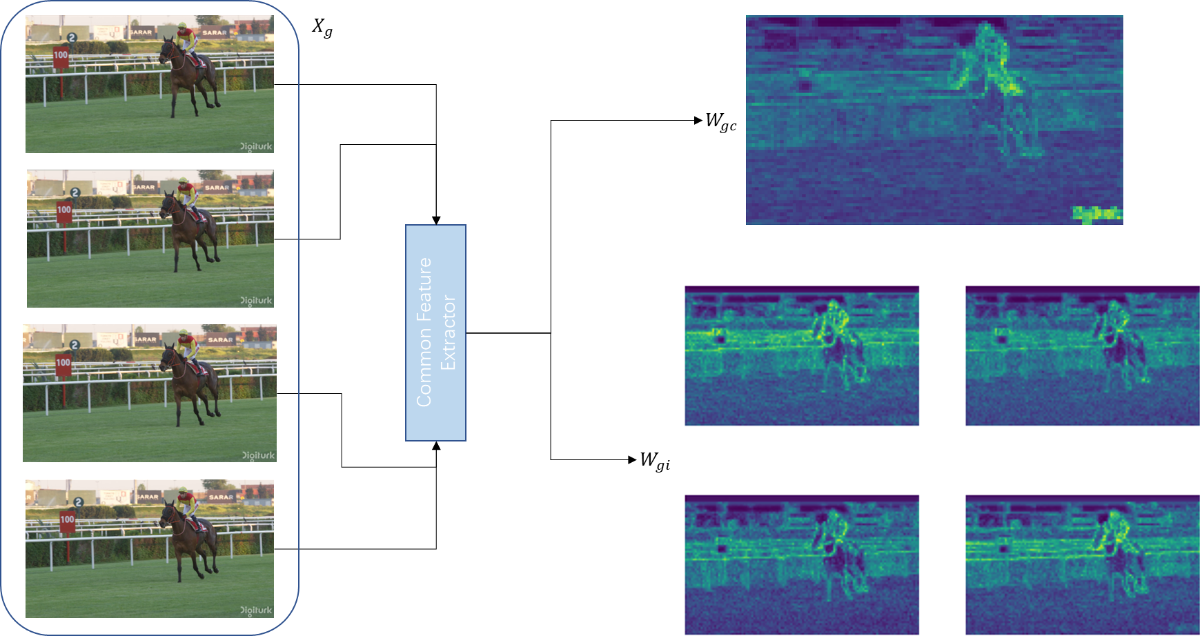}}
  \caption{Illustration of model division for common features $W_{gc}$ and individual features $W_{gi}$. The size of the GOP is set to 4. Modules and encoding processes before CFE are emitted for simplicity.\label{fig5}}
  \end{figure*}

\subsection{The framework of MDVSC}
The proposed MDVSC framework is presented in Fig.~\ref{fig1}. The input of the system is a stack of video frames called a GOP $X_g$. Before encoding the input GOP directly, the system first transforms them into latent space and forms latent representation $L_g$. This reduces the dimension of the data and speeds up computation when the spatial dimension is enormous. It also transforms the original data distribution from visible color space, such as RGB, to a latent space. It makes it easier for the JSCC encoder to extract semantic features further. 

After that, JSCC Encoder extracts semantic features from the latent representation $L_g$ and outputs semantic feature maps $Y_g$. Then, all feature maps will be fed into the common feature extractor (CFE). It will extract common features and individual features in this GOP and generate one common feature map $W_{gc}$ along with $N$ individual feature maps $W_{gi}$ for all frames. Combining these feature maps together is collectively referred to as $W_g$. This process can be formulated as
\begin{equation}
    W_{gi} = Y_g - W_{gc} ~\text{with}~W_{gc} = f_c(Y_g),
\end{equation}
\begin{equation}
    W_{g} = \{W_{gi}, W_{gc}\},
\end{equation}
where $f_c(\cdot)$ denotes the function to generate common features.
Apart from that, an entropy model is applied to measure the information entropy of the symbol and restrict the data distribution of the CFE, which is 
\begin{equation}
    em = f_e(W_g|d_\alpha,\theta),
    \label{formula2}
\end{equation}
where $f_e(\cdot)$ is the probability mass function, $em$ denotes the likelihood and $d_\alpha$ means the assumed distribution of $W_g$, Gaussian or Laplace distribution, for example. And $\theta$ is the weight and bias parameter in the deep entropy model. Details will be presented in subsection \emph{II-D}. The last process at the end of the transmitter is variable length coding. Based on the $W_g$ and its entropy, some elements will be dropped to satisfy certain constraints like bandwidth cost. The output of variable length coding is denoted as $S_g$.

Symbols $S_g$ are distorted to noisy signal $\hat{S_g}$ through the wireless channel. This paper mainly considers the widely used additive white Gaussian noise (AWGN) channel. In which case, the transfer function is $\mathcal{W}(\cdot~;\sigma^2)$ where $\sigma^2$ is the power of the noise, i.e.,
\begin{equation}
    \hat{S_g} = \mathcal{W}(S_g~;\sigma^2) = s_n + e_n,
\end{equation}
where $e_n$ is the vector of the Gaussian Noise and it satisfy $\sigma^2=\sum_1^n {e_n}^2$.
At the end of the receiver, the JSCC decoder inverses the process of the JSCC encoder and outputs latent representation $\hat{Y_g}$. Finally, the latent inversion module transforms $\hat{Y_g}$ back to its original distribution space and generates recovered GOP $\hat{X_g}$ directly. It should be noted that this framework assigns more computing burden to the transmitter, making it more practical to deploy the receiver to some computing resource-limited communication nodes. To sum up, the whole link of the MDVSC can be formulated as
\begin{equation}
\begin{split}
    X_g\xrightarrow{f_a(\cdot)}L_g\xrightarrow{g_a(\cdot)}Y_g\xrightarrow{f_c(\cdot)}W_g\xrightarrow{f_v(\cdot|em)}S_g\xrightarrow{\mathcal{W}(\cdot~;\sigma^2)}\hat{S_g},\\
    \hat{S_g}\xrightarrow{g_s(\cdot|em)}\hat{Y_g}\xrightarrow{f_s(\cdot)}\hat{X_g},~~~~~~~~~~~~~~~~~~~~~&
\end{split}
\end{equation}
where $f_a(\cdot)$ and $g_a(\cdot)$ means latent transform and JSCC encoding, $f_s(\cdot)$, $g_s(\cdot|em)$ denote the inverse function of the former. 

\subsection{Model division video transmission}
Semantic features that are similar across different frames in a GOP are called common features, and semantic features that are unique to each frame are called individual features. This concept is illustrated in Fig.~\ref{fig4}. Two consecutive video frames are encoded, and their feature maps are obtained. Then, one feature map is subtracted from the other, and the result is visualized. In Fig.~\ref{fig4}, the logo in the lower right corner of the original frame is a common feature. It appears in all frames of this GOP and is captured by the JSCC encoder. These common features do not need to be sent multiple times for video communication tasks. In fact, after subtraction, the logo’s semantic information disappears from the feature maps, which means that it can be transmitted only once, and channel bandwidth can be saved.

Therefore, a model division video semantic transmission scheme is proposed based on the idea above. Fig.~\ref{fig5} shows this scheme. The original GOP $X_g$ is latent transformed, JSCC encoded, and then fed into the CFE. The CFE extracts the common feature $W_{gc}$ and the individual features $W_{gi}$ from the GOP. The common feature $W_{gc}$ contains the shared semantic information of the GOP, such as the logo in the lower right corner, the grassland texture, the background, and the horse racer. It is smoother and more average than the original GOP. In contrast, the individual features $W_{gi}$ are sharper and more specific. They contain the personalized semantic information of each frame and remove shared duplicate features, such as the horseshoes’ position and the horse’s outline. The static logo is also removed and not included in the individual features.

\subsection{Entropy-based variable length coding}
An explicit code length control scheme is proposed to maximize transmission performance while satisfying communication limitations. It is based on the information entropy to measure which data should be preserved. Therefore, it is essential to design an entropy model to estimate the probability distribution of the input semantic features. Similar to the work of \cite{Minnen2018JointAA}, the semantic features $W_g$ are variationally modeled as the Gaussian distribution, where each scalar $w_{g,i}$ is of varying distribution parameters. In this paper, the mean of the distribution is fixed to zero to restrict the distribution of $W_g$. Therefore, the entropy of each scalar $w_{g,i}$ can be computed as
\begin{equation}
    r_{g,i}=-\log P_{\Bar{w}_{g,i}|\Bar{z}_{g,i}}(\Bar{w}_{g,i}|\Bar{z}_{g,i}).
    \label{formula1}
\end{equation}
where $\Bar{w}_{g,i}=\lfloor w_{g,i} \rceil$ is the quantization to the input data. Similarly, $\Bar{z}_{g,i} =\lfloor z_{g,i} \rceil=\lfloor f_\text{hpe}(w_{g,i}) \rceil$, and $f_\text{hpe}(\cdot)$ denotes the hyperprior encoding function for obtaining the distribution parameter. In some former works, $w_{g,i}$ is rounded to an integer for the convenience of entropy coding. However, in the MDVSC, as the entropy coding is deprecated, this operation is used for entropy estimation and improving the robustness of the JSCC codec. The quantization method is implemented by using the soft quantization proposed in the \cite{Ball2016EndtoendOI}. Therefore, $W_g$ is further modeled as
\begin{equation}
    \Tilde{W}_g=W_g+O~\text{with}~o_i \sim \mathcal{U}(-\frac{1}{2}, \frac{1}{2}),
\end{equation}
therefore $\Tilde{w}_{g,i}$ is variationally modeled as a Gaussian distribution with learned variance $\Tilde{\sigma}_{g,i}$, i.e.,
\begin{equation}
\begin{split}
    p_{\Tilde{W_g}|\Tilde{Z_g}}(\Tilde{W_g}|\Tilde{Z_g})=\prod_i P_{\Tilde{w}_{g,i}|\Tilde{z}_{g,i}}(\Tilde{w}_{g,i}|\Tilde{z}_{g,i})\\
    =\prod_i (\mathcal{N}(0, \Tilde{\sigma}_{g,i})*\mathcal{U}(-\frac{1}{2}, \frac{1}{2}))(\Tilde{w}_{g,i}),\\
    \text{with}~\Tilde{\sigma}_{g,i}=f_\text{hpd}(\Tilde{z}_{g,i})~~~~~~~~~&
\end{split}
\end{equation}
where $*$ is the convolutional operation. $\mathcal{N}(0, \Tilde{\sigma}_{g,i})$ is the Gaussian distribution where the mean is fixed to zero and variance $\Tilde{\sigma}_{g,i}$ is calculated from the hyperprior decoding function $f_\text{hpd}(\cdot)$. Herein, with the preset Gaussian distribution along with its parameter mean and variance, the entropy of each scalar can be calculated according to Eq.~\ref{formula1}.

\begin{figure}
  \centerline{\includegraphics[width=3.4in]{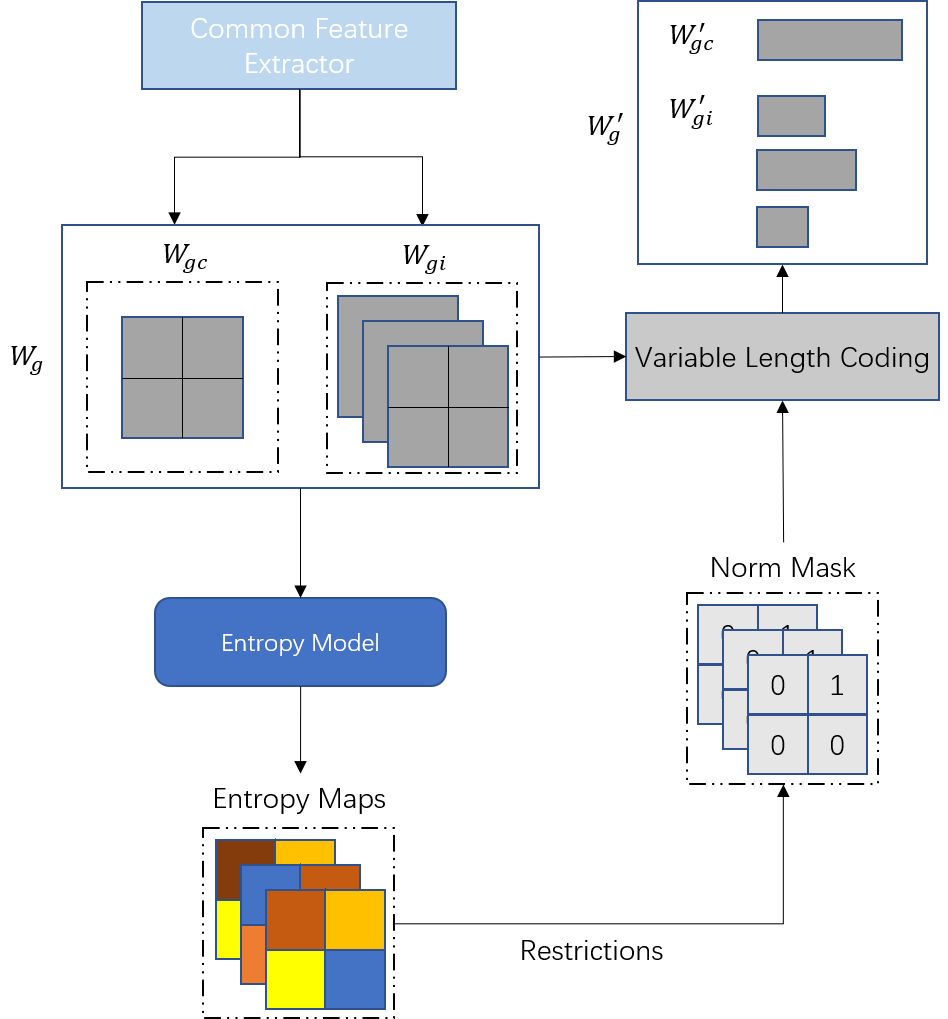}}
  \caption{The illustration of the entropy-based variable length coding scheme. For the scalar in both the common feature $W_{gc}$ and the individual features $W_{gi}$, entropy is assumed to be the weight for measuring the importance of every element in feature maps. Based on this and resource limitation, a norm mask can be calculated to decide which semantic feature should be preserved.\label{fig3}}
  \end{figure}

Based on the derivation, the entropy-based variable length coding is proposed. The illustration is shown in Fig.~\ref{fig3}. Following the output $Y_g$, which is wrapped by the JSCC Encoder, the CFE analyses it and outputs $W_g$. It contains a common feature map $W_{gc}$ and several individual feature maps $W_{gi}$. They will be fed into the entropy model and get entropy maps, which measure the importance of each element in $W_g$. Then, according to the restrictions, which can be set manually and explicitly, a norm mask will be calculated correspondingly. After that, the original feature maps and the norm mask can be sent into the variable length coding module and get variable length codes $W_g^{~'}$. It is composed of common feature vector $W_{gc}^{~'}$ and individual feature vectors $W_{gi}^{~'}$, both of which have controllable length. Relative experiments are shown in section \emph{IV-C-3}. Finally, $W_g^{~'}$ gets transformed to $S_g$ and sent through the wireless channel.

\begin{figure*}
  \centerline{\includegraphics[width=6.8in]{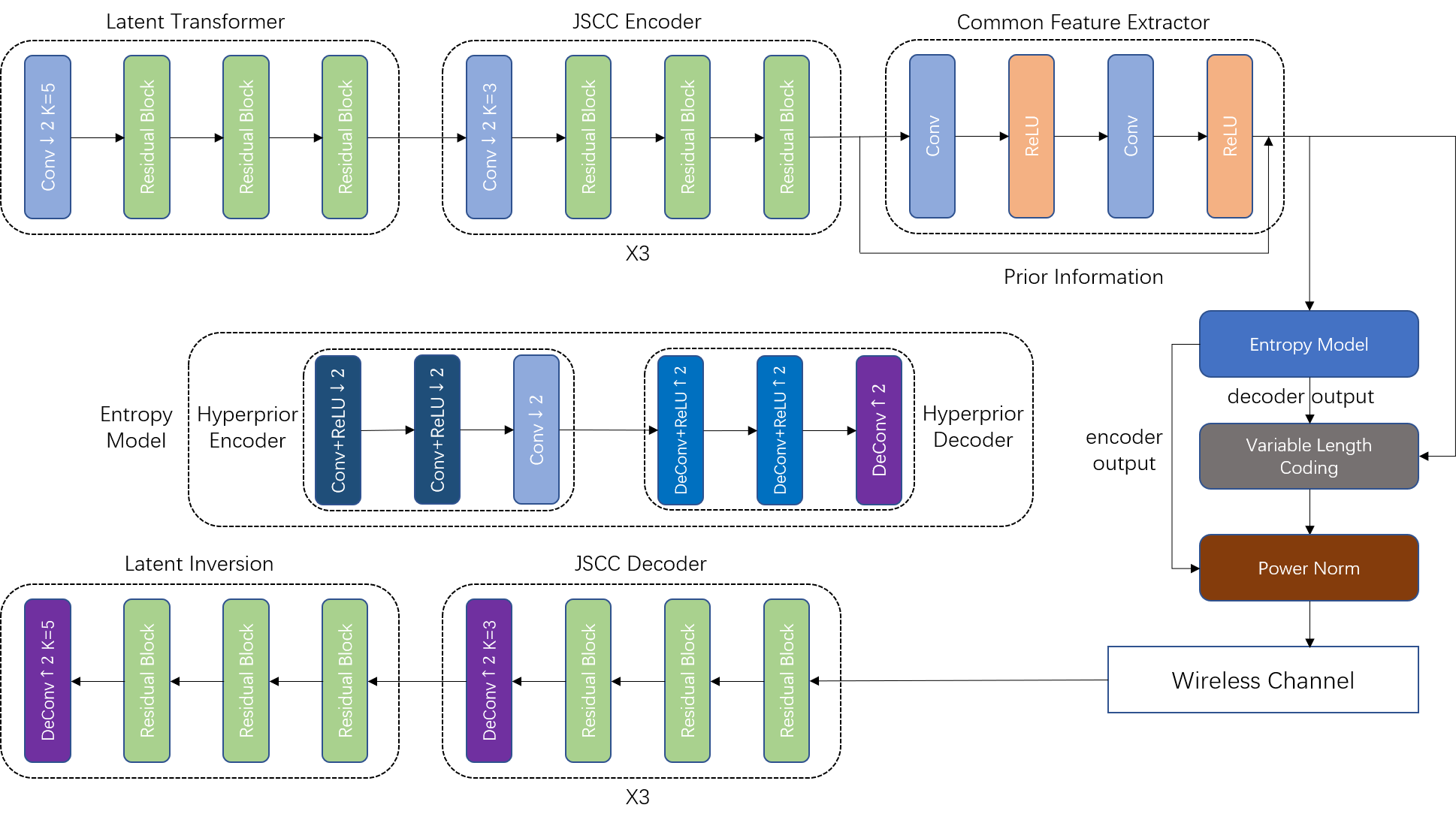}}
  \caption{Network architectures of the MDVSC. $\uparrow$ 2 or $\downarrow$ 2 indicates upscaling or downscaling with a stride of 2. K=N means a convolution layer with N$\times$N filters. Conv or Deconv without an arrow indicates that there is no dimensional change.\label{fig2}}
  \end{figure*}
  
\begin{figure*}
  \centerline{\includegraphics[width=6.8in]{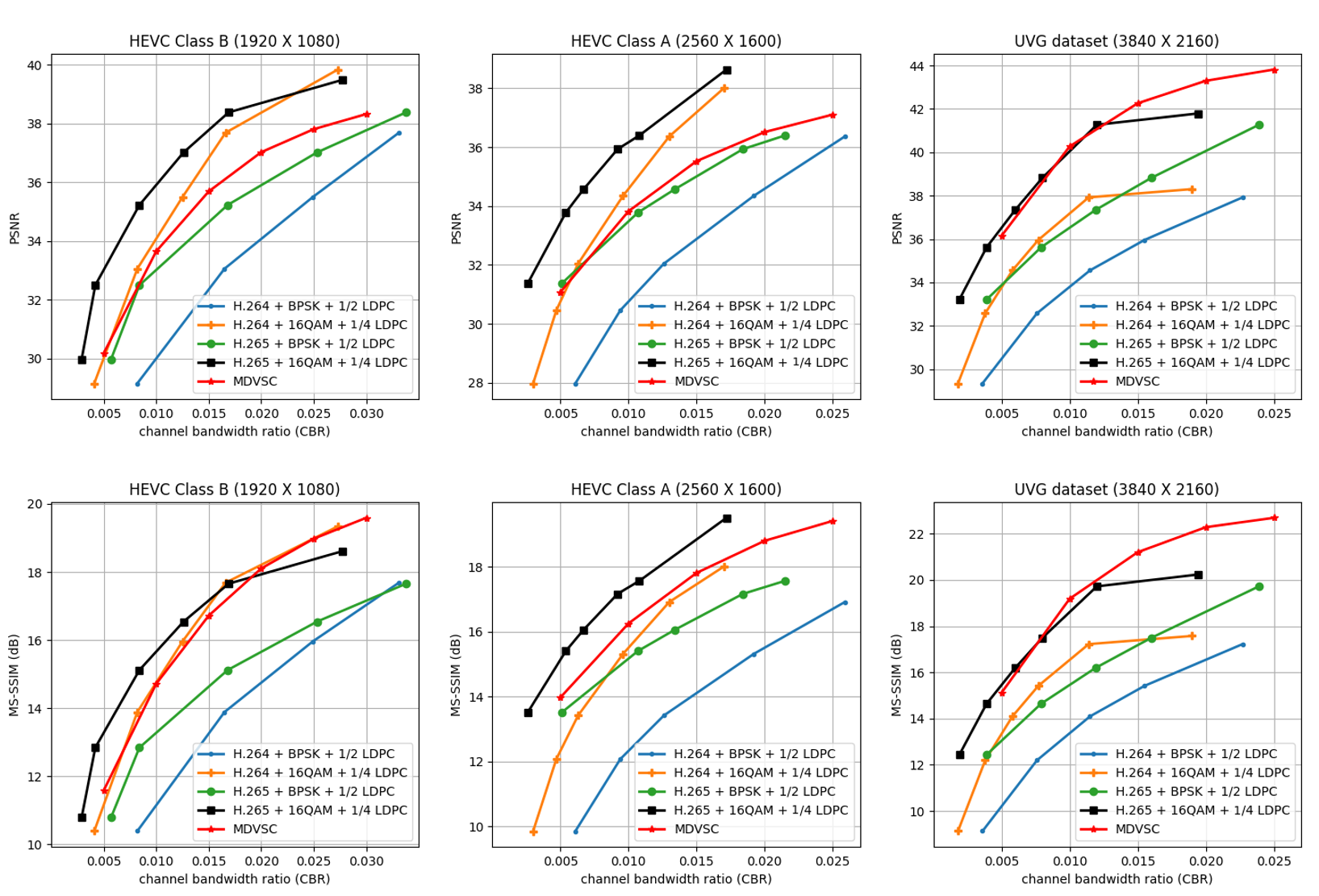}}
  \caption{PSNR and MS-SSIM performance versus the channel bandwidth ratio (CBR) over the AWGN channel at SNR = 15dB. The value of MS-SSIM is converted to dB with Eq.~\ref{formula13}. In addition, it can be found that the red data point of the MDVSC coincides with the coordinate axis and is distributed at equal intervals even for different videos, indicating that it has precise code length control capability. In comparison, the traditional scheme can only approximate the given CBR.\label{fig6}}
  \end{figure*}

\subsection{Optimization goal}
The optimization goal of the MDVSC is to improve video reconstruction quality as much as possible under limited communication resources such as the CBR. Given that the MDVSC takes the GOP as the input, the loss function of the $n$th GOP can be formulated as 
\begin{equation}
    Loss = \lambda \cdot K_n + D_n = \lambda \cdot \sum_{t = 1}^{N} k_t + D_n,
    \label{formula3}
\end{equation}
where $\lambda$ controls the trade-off between the CBR $K_n$ and the distortion $D_n$. Because our system learns the overall channel bandwidth allocation across sequential frames for a GOP, the $k_t$ for each frame is different. During training, the CBR term $K_n$ is replaced with another metric entropy, mentioned in Eq.~\ref{formula2}, for the convergence of the entropy and hyperprior models. Therefore, the formula Eq.~\ref{formula3} works out to be
\begin{equation}
    Loss = -\lambda \cdot \log_2(f_e(W_g|d_\alpha,\theta)) + D_n.
\end{equation}
It is worth noting that in our system, the trade-off between performance and communication cost does not rely on the hyperparameter $\lambda$. Therefore, it is not changeable and fixed to a certain number. 

\section{Network architectures}
The details of the proposed MDVSC are illustrated in Fig.~\ref{fig2}. The whole system is asymmetric and assigns more computing burden to the transmitter. In the following subsections, details about each module will be presented.

\subsection{Transmitter}
\subsubsection{Latent Transformer}
Video semantic communication in the visible space demands high computation capability from the transmitter and receiver. As video data dimension increases, real-time video communication becomes more challenging. To address this issue in a certain extent, the latent transformer is introduced to map video frames from the original color space to the latent space. A down-sampling convolution layer and three residual layers are used to transform the original data into a latent space that is more suitable for video communications and lowers computing complexity.

\subsubsection{JSCC Encoder}
The JSCC encoder in this paper consists of three consecutive blocks. Each block has a down-sampling convolution layer and three residual layers. This module encodes the latent representations from the latent transformer and adapts to the channel conditions. It produces semantic feature maps resilient to channel noise and sends them to the next module.

\subsubsection{Common Feature Extractor}
MDVSC aims to extract the common and individual features in a GOP. This is done by a CFE that has two convolution and activation layers. The common feature is the output of the CFE, and the individual feature is the remaining semantic information. To make it easier to converge, prior information is given as a residual connection. In our training, the mean value of the whole GOP is used as the prior information. In this way, the CFE only has to learn the difference between the common feature and the mean semantic feature, which speeds up the convergence of this module.

\subsubsection{Entropy Model And Variable Length Coding}
The entropy model consists of a hyperprior encoder and a hyperprior decoder. In some previous works, the hyperprior encoder’s output is used as the side information that is sent through the wireless channel to improve the frame reconstruction quality. In our system, this side information is used for variable length decoding. As stated before, the feature maps are assumed to have a Gaussian distribution, and the hyperprior decoder computes their variance. Therefore the entropy maps for $W_{gc}$ and $W_{gi}$ are obtained and sent to the variable length coding module to produce variable length codewords. These codewords are then transmitted through the wireless channel after the power norm module.

\subsection{Channel}
The wireless channel module is a non-trainable layer that links the transmitter and the receiver. The AWGN channel is mainly used in this paper, but other channel models can also be embedded by changing this layer. The wireless channel’s effect is determined by a parameter called signal-to-noise ratio (SNR). It is represented as
\begin{equation}
    SNR=10\log_{10}(\frac{\Tilde{p}}{n}),~\text{with}~\Tilde{p}=f_{pn}(p)=1,
\end{equation}
where $p$ and $n$ are the power of the signal and noise, respectively. $f_{pn}(\cdot)$ denotes the power norm function to normalize the signal's power to 1.
 
\subsection{Receiver}
\subsubsection{JSCC Decoder}
The JSCC decoder in this paper has the same structure as the JSCC encoder, with three blocks of one transpose convolution layer and three residual layers. It decodes the codewords $\hat{S_g}$ into the latent representation $\hat{Y_g}$. 

\subsubsection{Latent Inversion}
At the end of the MDVSC, the latent inversion module maps the latent representation $\hat{Y_g}$ back to the original color space, which is visible to humans. This completes the transmission of this GOP. 

\section{Experiments}
Some experiments and results for MDVSC are presented in this section. Subsection \emph{A} will introduce the experiment setup and metrics. The overall communication ability of MDVSC is compared with the traditional separated scheme in different CBRs and SNRs in subsection \emph{B}. The uniqueness of MDVSC, including video jitter, jump frames, and variable coding, is shown by extended experiments that are conducted in subsection \emph{C}. The effectiveness of the proposed method is demonstrated by the ablation study that is conducted in the subsection \emph{D}.
\subsection{Experimental Setup}
\subsubsection{Datasets}
Our MDVSC is trained with the Vimeo-90k dataset \cite{Xue2017VideoEW}, which has 89800 video clips with various scenes and actions. Each video clip has 7 frames in a sequence. We set the GOP size to 6 and randomly crop the frames to 256 $\times$ 256 pixels during training. After training, The performance is evaluated with the HEVC test dataset \cite{Bossen2010CommonTC} and UVG dataset \cite{Mercat2020UVGD5}. The following subsets are selected: Class A (2560 $\times$ 1440), Class B (1920 $\times$ 1080), and UVG (3840 $\times$ 2160) because they have higher resolution that is more common in actual use, and they challenge the communication systems more.

\begin{figure*}
  \centerline{\includegraphics[width=6.8in]{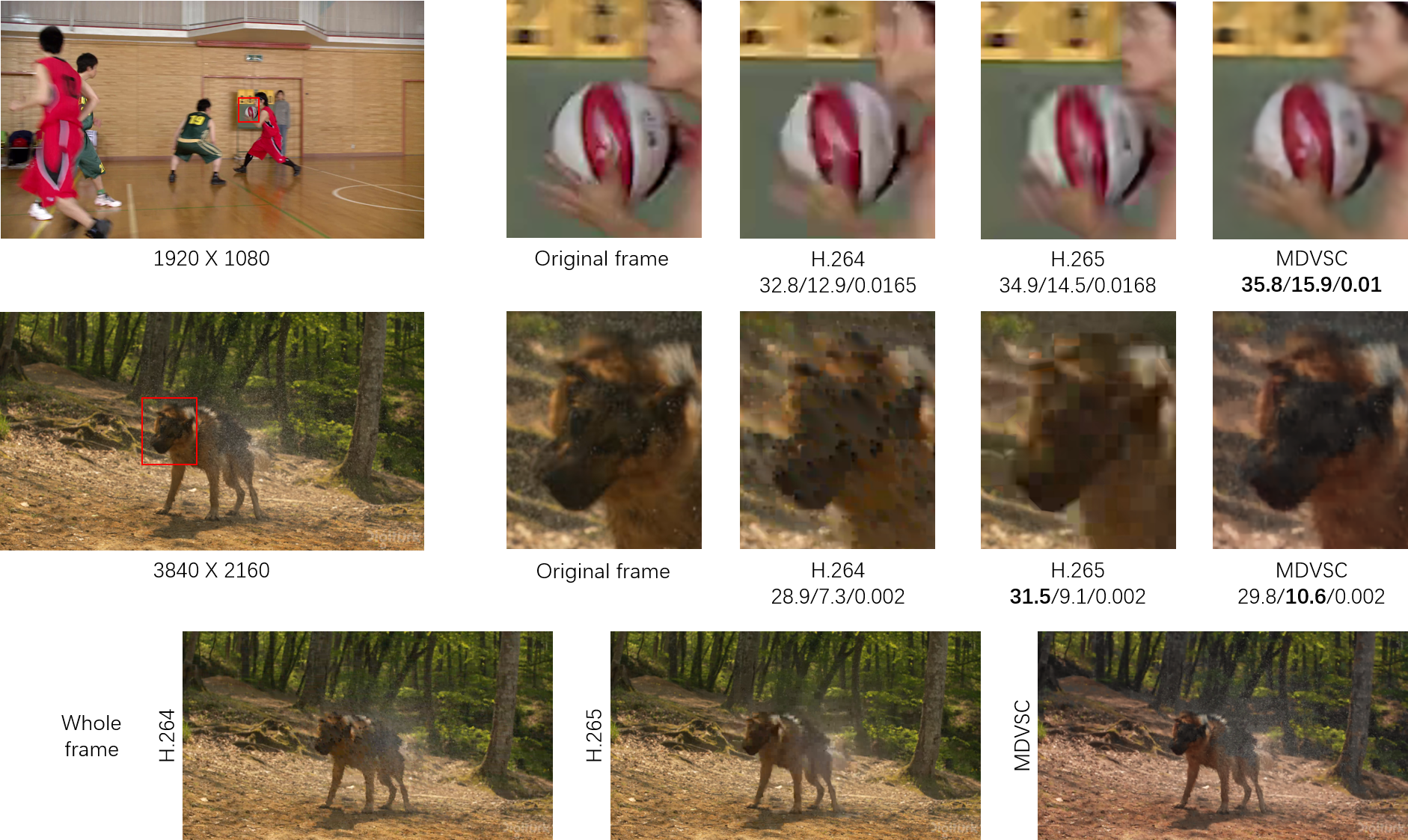}}
  \caption{Examples of visual comparison. For the first two lines, the first column shows the original frame. The second column shows the cropped patch in the original frame. The third to fifth columns offer the reconstructed frames by using different schemes shown in Fig~\ref{fig6}. The subtitle is PSNR/MS-SSIM/CBR, respectively. Channel coding and modulation for H.264/H.265 are 1/2 LDPC and BPSK. The first line shows their performance in the typical compression rate (CBR), while the second line shows the performance at a very high compression rate. The third line shows the whole frame of the corresponding cropped patch in the second line.\label{fig7}}
  \end{figure*}

\subsubsection{Implementation Details}
In all experiments, the channel dimension for latent representations and JSCC codewords is 128. The MDVSC is trained in terms of the mean square error (MSE) and tested under the PSNR, or MS-SSIM \cite{Wang2003MultiscaleSS} for perceptual quality. These three metrics can be calculated as follows: 
\begin{equation}
    MSE(X, Y) = \frac{1}{mn}\sum_{i=0}^{m-1}\sum_{j=0}^{n-1}[X(i, j)- Y(i, j)],
\end{equation}
where $m$ and $n$ denote the number of pixels horizontally and vertically.
\begin{equation}
    PSNR(X, Y) = 10 \cdot \log_{10}(\frac{1}{MSE}),
\end{equation}
\begin{equation}
\begin{split}
    MS-SSIM(X, Y) = \\
    [\mathcal{L}_M(X, Y)]^{\alpha_M}\cdot\prod_{j=1}^M[\mathcal{C}_j(X, Y)\cdot\mathcal{S}_j(X, &Y)]^{\alpha j},
\end{split}
\end{equation}
where M means different dimensions, including luminance, contrast, and structure. Therefore, the PSNR metric measures the accuracy of the frame reconstruction, while the MS-SSIM metric evaluates the perceptual quality of the recovered frame. The hyperparameter $\lambda$ is set to 8192 for all training processes because the rate-distortion (RD) tradeoff in our system does not depend on it. The learning rate for training is $10^{-4}$ at the beginning, and cosine annealing algorithm \cite{Loshchilov2016SGDRSG} is used for decay. The mini-batch size is 32, and the whole MDVSC model is trained on two RTX A6000 GPUs.

\begin{figure*}
  \centerline{\includegraphics[width=6.8in]{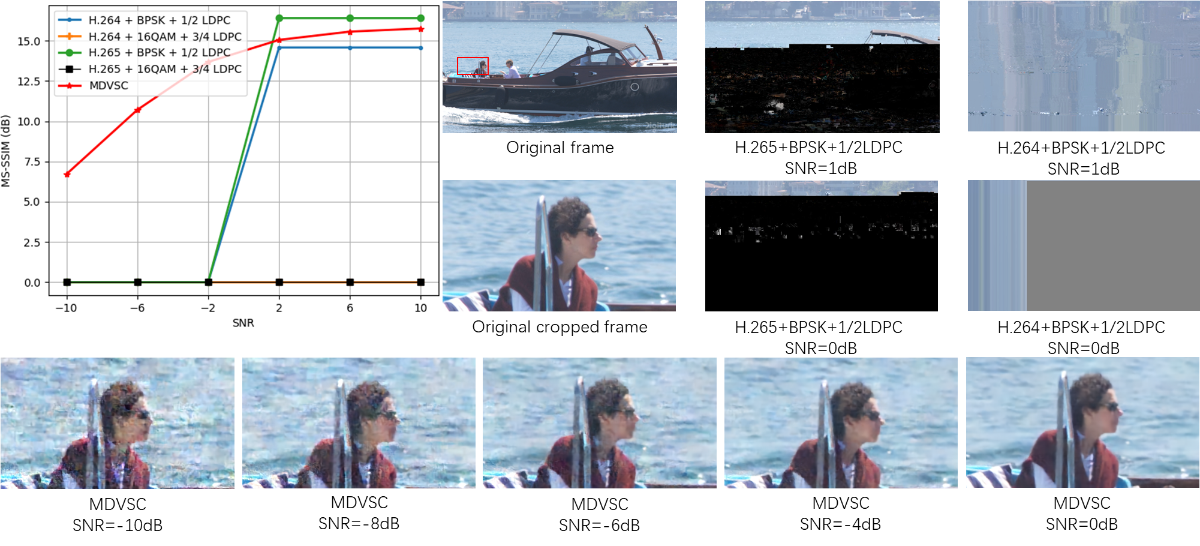}}
  \caption{MS-SSIM performance versus the change of channel SNR over AWGN channel. Even under severe channel conditions, the MDVSC can still work properly and hardly suffer from the cliff effect. \label{fig8}}
  \end{figure*}

\subsubsection{Comparison Schemes}
The MDVSC is compared with the classical video coding transmission schemes, including source coding H.264 \cite{Wiegand2003OverviewOT} and H.265 \cite{Sullivan2012OverviewOT}, channel coding LDPC \cite{Richardson2018DesignOL}, modulation BPSK and 16QAM for noise resistant and transmission efficiency respectively. The FFmpeg preset is set to placebo. 

\subsection{Result}
The PSNR and MS-SSIM performance curves are shown in Fig.~\ref{fig6}. The SNR is set to 15dB to ensure the performance of the 16QAM. Since MS-SSIM yields values that are higher than 0.9 mostly, it is converted to dB to improve the legibility \cite{Ball2018VariationalIC}, i.e.,
\begin{equation}
     MS-SSIM~(dB) = -10\log_{10}(1 - d), \text{with}~d\in[0, 1],
    \label{formula13}
\end{equation}
where $d$ is the MS-SSIM value that ranges from 0 to 1. The MDVSC performs better than the noise-robust combination BPSK + 1/2 LDPC, generally for both H.264 and H.265, under the PSNR metric. The transmission efficiency combination 16QAM + 1/4 LDPC performs better than the MDVSC. However, for the higher resolution scene of 3840 $\times$ 2160, the MDVSC can surpass the transmission efficiency combination for H.264 and closely match that of H.265.

The results are different under the MS-SSIM metric, which assesses perceptual quality that is more aligned with human feelings. The MDVSC outperforms the transmission efficiency combination for H.264 and reduces the gap with that of H.265. Also, the MDVSC shows significant coding gain as that of the noise-robust combination for both H.264/H.265 series.

Moreover, it is easy to see that the CBR parameter of the MDVSC in Fig.~\ref{fig6} is always equally spaced and varies from 0.005 to 0.03 at the same step for different videos. It means that the MDVSC can control CBR accurately to match the threshold. In contrast, other video communication schemes can only approximate the given CBR threshold.
  
Fig.~\ref{fig7} shows the reconstructed frames for different schemes. The channel coding and modulation scheme for H.264/H.265 are 1/2 LDPC and BPSK. The CBR is 0.01 for the first line and 0.002 for the second line to show their performance at different compression rates. For the first line, It can be seen that the MDVSC produces more sharp and clear frames, especially for the logo on the basketball. This shows the competitive performance of the MDVSC at the typical compression rate. The CBR is very low for the second line to simulate severe communication bandwidth. It can be seen that H.264/H.265 have noticeable artifacts and block effects while the MDVSC generates a cleaner texture. However, the MDVSC also changes the hue of the frame, which is more apparent in the third line where the whole frame is shown. For the MDVSC, when the CBR is too low to transmit enough data, it will drop color-related data to preserve the more perceptually important object to humans, such as the edge and the outline. This explains why the PSNR and MS-SSIM metrics give different conclusions to the recovery in the second line.

Fig.~\ref{fig8} shows the MS-SSIM and visualization results versus the channel SNR, where the CBR constraint is 0.01. PSNR results are omitted for their similar trend with the MS-SSIM. It is noted that the MDVSC is trained only once under the SNR 10dB and tested under different SNRs. It can be seen that the MDVSC works well even under very harsh channel conditions. Its performance shows a gradual decline trend with the decrease of SNR. In contrast, traditional schemes experience significant cliff-effect and fail to work properly when the SNR is lower than a certain threshold. Bit errors not corrected by LDPC coding make it difficult for traditional video decoders to recover video frames correctly, especially when these errors are control bits.

Moreover, in the downlink video transmission scenario, the receiver's decoding time is tested. For the video with 3840 × 2160 resolution, a 13900K CPU and RTX A6000 GPU are used to run the decoder. The test results show that the average decoding time for one frame is about 8ms.

\subsection{Extended experiments}
\subsubsection{Video jitter experiments}
 \begin{figure*}
  \centerline{\includegraphics[width=7in]{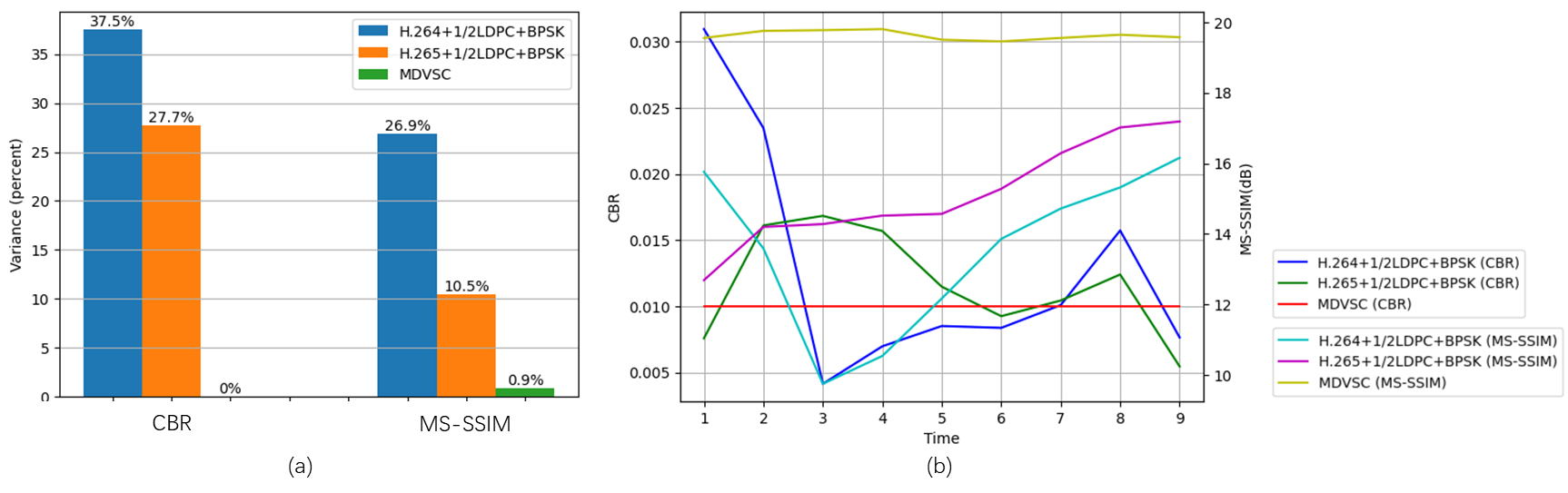}}
  \caption{Video CBR and performance variance test results. Subgraph (a) shows the overall CBR and MS-SSIM variance according to their mean value. Subgraph (b) shows one of the videos' real-time CBR and performance fluctuations of the traditional and semantic communication schemes. \label{fig15}}
  \end{figure*}

It is shown that the code length can be adjusted accurately and flexibly by our MDVSC, which can reduce the variation of transmission rate and enhance performance. This is demonstrated by comparing the MDVSC with the separated coding schemes on different videos from the test dataset. The CBR is set to 0.01, and the results are presented in Fig.~\ref{fig15}. Subgraph (a) displays the overall mean CBR and MS-SSIM variance at HEVC and UVG test sequences. It is observed that the conventional communication schemes suffered from inevitable fluctuations in bandwidth and video quality. In contrast, the bandwidth is controlled more accurately, and MDVSC maintains a more stable video quality. Subgraph (b) illustrates one of the videos' real-time variations of CBR and MS-SSIM. Since the code length cannot be adjusted precisely by the separated scheme, it might experience a jitter when the CBR exceeds the limit while the local cache is empty.

\subsubsection{Jump frames experiments}

\begin{figure*}[ht]
  \centerline{\includegraphics[width=6.8in]{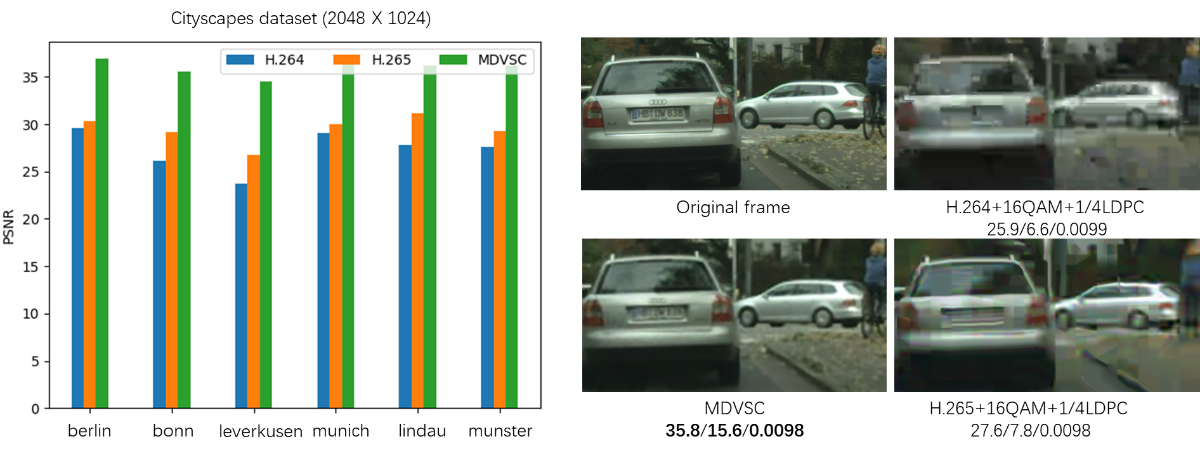}}
  \caption{Test results at different scenes are shown on the left. 16QAM + 1/4 LDPC is used to ensure the transmission efficiency of the separated coding scheme. It shows that the MDVSC can achieve much better coding gain than other schemes when facing jump frames. The visualization of jump frames reconstructed from different communication schemes is shown on the right. The subtitles of different frames are PSNR/MS-SSIM/CBR, respectively. \label{fig9}}
  \end{figure*}
  
 \begin{figure}
  \centerline{\includegraphics[width=3.4in]{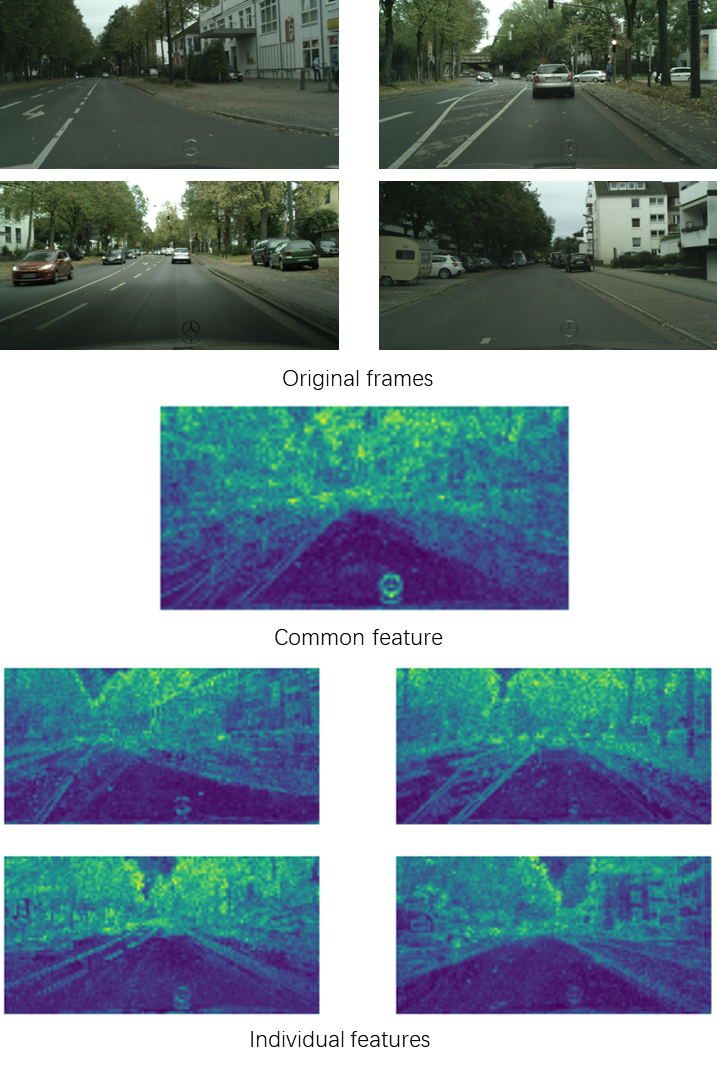}}
  \caption{The visualization of common features and individual features for jump frames. Individual features are similar to their corresponding frame, while the common feature is more complicated and fuses more semantic features into one feature map. \label{fig10}}
  \end{figure}

To further investigate the mechanism and potential of the MDVSC, a series of special videos with jump frames are designed based on the Cityscapes Dataset \cite{Cordts2016TheCD}. Most of the frames in these videos are not continuous at the time domain and nearly have no visualized common object. It has to be noted that the MDVSC does not get fine-tuned for this class of videos. It is only pre-trained on the Vimeo-90k dataset, whose video frames are continuous at the time domain. When setting the SNR to 15dB and using 16QAM + 1/4 LDPC coding, test results and visualization results are shown in Fig.~\ref{fig9}. The CBR limitation is set to 0.01. It is clear that traditional source and channel separate coding schemes cannot properly cope with jump frames. A possible reason is that the MEMC system can hardly find an appropriate reference in the intra-frame for the current predictive frame, making the bits needed for each frame extremely increased. However, it is found that the MDVSC can still work properly, even for jump frames. To explain that, the visualization of the common feature and individual features is shown in Fig.~\ref{fig10}. Individual features for each frame are similar to their original frames, which is the same as before, while the common feature is quite different from any of the individual features. From where we stand, CFE extracts some more abstract features across video frames and tries to find similar information, which is in higher dimension semantic space instead of the visual space like RGB. It makes the MDVSC robust and suitable for more application scenarios.

\subsubsection{Entropy-based variable length coding ratio}
\begin{figure}
  \centerline{\includegraphics[width=3.4in]{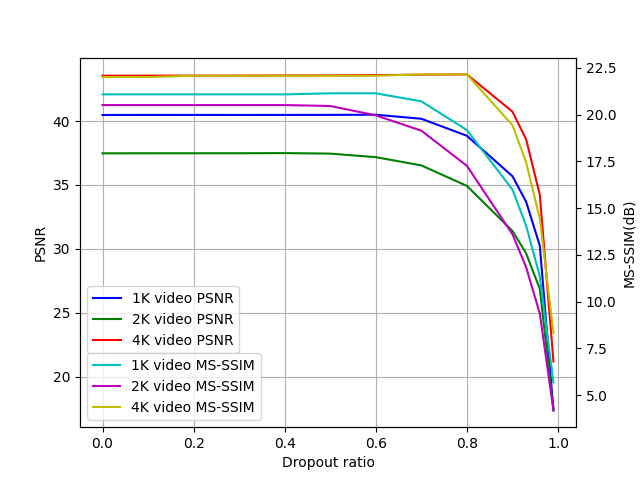}}
  \caption{Test results of various dropout ratios. It can be seen that nearly half of the data can be discarded without affecting performance. It proves the validity of the entropy-based variable length coding scheme. \label{fig11}}
  \end{figure}

In MDVSC, an entropy-based variable length coding scheme is proposed to control data amount explicitly and precisely, flattening the delay jitter and improving communication link performance. There is a threshold to control the percent of data to be dropped, namely the dropout ratio. It is a floating index that ranges from 0 to 1, and 0.5 means drop half data of the semantic feature, for example. In our system, there are two parts of information to be clipped, i.e., common features and individual features. Firstly, their dropout ratios are set to be the same, and test results are shown in Fig.~\ref{fig11}. It can be figured out that nearly half of the semantic features with low information entropy can be discarded without descending the performance significantly, indicating that tons of data inside the semantic feature map is redundant and is of little use to the reconstruction of the video frame. Instead, this redundancy lowers the transmission efficiency of semantic communication. Therefore, our entropy-based variable length coding can help to recognize useful semantic features, namely information with high entropy, and drop out other data to improve communication performance as much as possible. Besides, it is found that 4K video tends to have a higher dropout ratio turning point, which indicates that these video data have more correlation at semantic feature space and are robust to data missing. 

\begin{figure}
  \centerline{\includegraphics[width=3.4in]{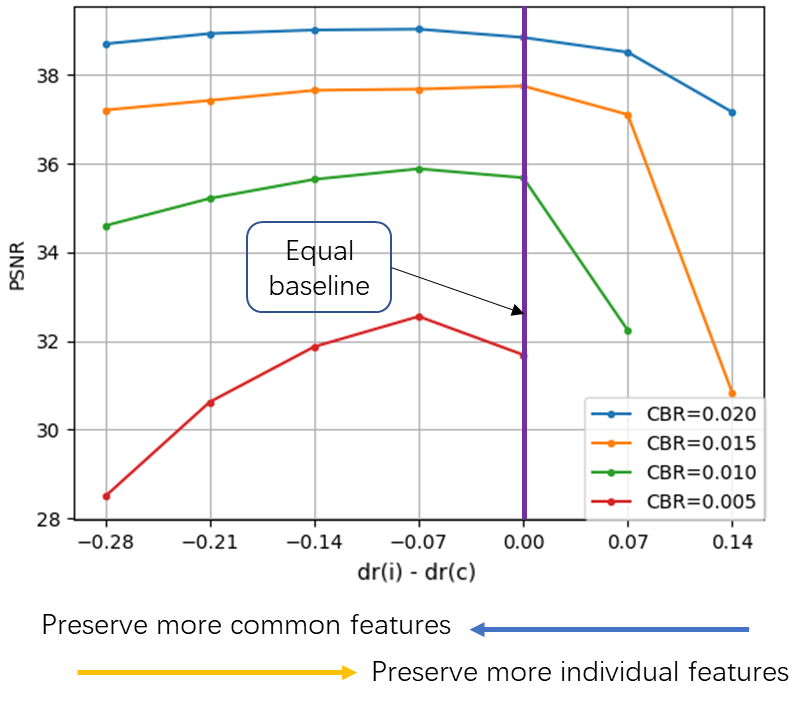}}
  \caption{Test results for changing the balance between common and individual features dropout ratio. From left to right means preserving more individual features and vice versa. The vertical line is the baseline where $dr(c)=dr(i)$. In general, preserving a bit more common feature can help improve performance, especially when channel limitation is strict. However, with the continual loss of the individual features, the performance gain will disappear, indicating the importance of balancing the common and individual features.\label{fig12}}
  \end{figure}

Secondly, considering that common features and individual features are of different information entropy, and generally, common features have higher entropy, it is not reasonable to set the same dropout ratio for both of them. Therefore, to meet the limitation of the CBR while reducing the loss of information entropy, we conduct a test of various dropout ratios for common features and individual features versus reconstruction quality. It has to be paid attention that to meet the bound of a fixed CBR precisely, the dropout ratio cannot change smoothly. The trade-off between them has to satisfy 
\begin{equation}
    baseline = [dr(c) \pm (t \cdot n_{GOP})] + [dr(i) \pm t],
\end{equation}
where $baseline$ is a fixed variable during one GOP transmission, $dr(c)$ and $dr(i)$ denote the drop ratio for the common feature and individual feature, respectively, $t$ is the trade-off weight and $n_{GOP}$ is the size of GOP. Based on the above, the test results are presented in Fig.~\ref{fig12}. The MS-SSIM metric is ignored for its similar trend with the PSNR. From the right to the left means preserving more common features and vice versa. The vertical line is the baseline where $dr(c)d=dr(i)$. The CBR of each fold line is fixed to a certain number. It can be found that common features are vital to frame reconstruction in nearly all channel limitations. When the channel restriction is loose, missing data on the individual features will not affect communication dramatically. Instead, more channel bandwidth for the common features can help slightly improve performance. It proves that individual features have more redundancy than common features. Nevertheless, the performance still descends if too many individual features are dropped. However, when the channel condition is strict, both features are greatly compressed, and the balance between them is important to the communication performance. For the line of CBR $=0.005$ in Fig~.\ref{fig12}, preserving more common features can improve performance for a while, but the loss of individual features will be dominant eventually. Therefore, allocating more bandwidth to the common features is helpful in improving end-to-end performance. 

\subsection{Ablation study}
\subsubsection{Common feature extractor}
\begin{figure}
  \centerline{\includegraphics[width=3.4in]{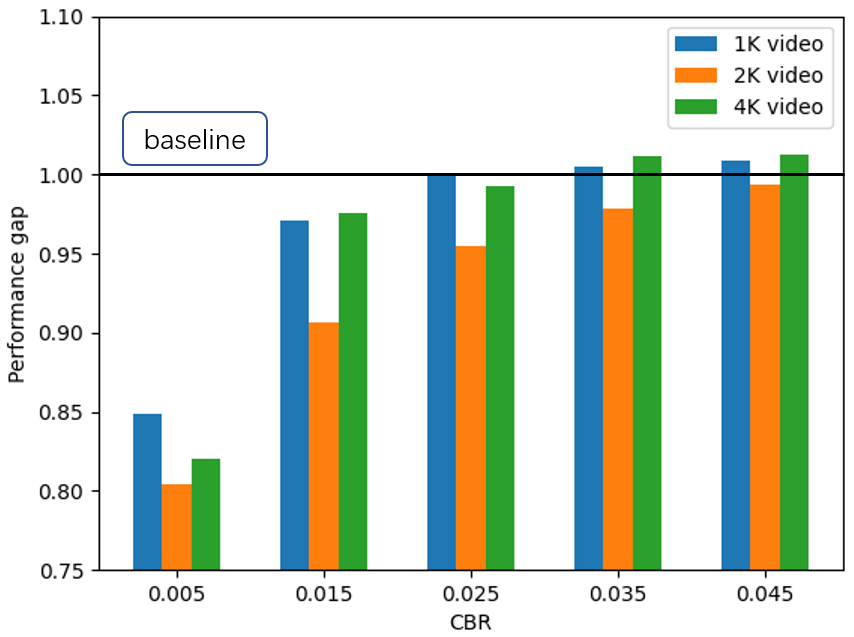}}
  \caption{Ablation study for the common feature extractor (CFE). It shows the performance gap with the baseline when the CFE gets removed. When the channel limitation is loose, the CFE damages the performance a bit for the imprecise common feature output. However, with the lower CBR, it starts to play a vital role in video transmission tasks for its ability to extract common features and reduce redundancy. \label{fig13}}
  \end{figure}

To quantify the function of the CFE, we remove it from the network and test its communication performance. The classical MDVSC is referred to here as the baseline. As the MDVSC is able to extract semantic features adaptively and form common features and individual features, so there is no need to retrain it if common features are lost. Test results are shown in Fig.~\ref{fig13}. When the channel bandwidth is enough to transmit more data, namely, the CBR is large enough, it can be found that the MDVSC's performance without CFE can slightly surpass the baseline. This can be explained by the fact that the CFE cannot extract common features quite accurately, which inevitably introduces noise for frame recovery and then reduces the baseline performance. It indicates that when channel bandwidth is enough, the architecture of CFE may not be necessary. However, with the descent of the channel capacity, there is not enough bandwidth for the MDVSC to transmit duplicate and redundant information. To meet the limitation, it has to drop more valuable information besides those repeated data, decreasing the performance. In this case, the CFE is critical in recognizing common features and sparing more channel resources for those valuable data. Therefore, the CFE is more significant for performance improvement under limited channel conditions.

\subsubsection{Variable length coding}
\begin{figure}
  \centerline{\includegraphics[width=3.4in]{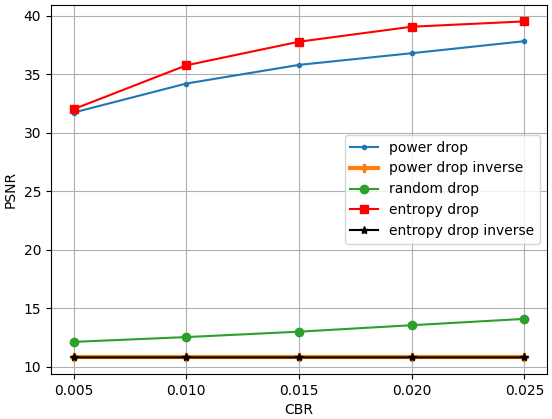}}
  \caption{Ablation study for entropy-based variable length coding. It is denoted as the entropy drop and is replaced with the power drop and random drop. The former drops data according to the power, while the latter drops randomly. The inverse version, where higher entropy or power data gets dropped, is also studied. It proves the effectiveness of the entropy-based variable length coding. \label{fig14}}
  \end{figure}

In order to figure out the function of the entropy-based variable length coding, it is replaced with the power-based variable length coding and the random-based variable length coding to view their performance gap. The former means weighing each data according to its power, while the latter drops data randomly. Besides, the inverse version, where data with higher entropy or a higher power will be dropped instead, is also studied. The classical MDVSC is referred to as the entropy drop. Test results are shown in Fig.~\ref{fig14}. It is clear that the inverse version of the power or the entropy drop scheme cannot work properly, making the MDVSC unable to transmit any valuable data. It proves that data with higher power or entropy is helpful to the transmission task. For the random drop scheme, its performance increases a little bit along with the loose of the channel limitation. However, its performance is still not sufficient to support high-quality communication. Entropy drop, namely the entropy-based variable length coding proposed in the MDVSC, performs better than the power drop scheme. When channel condition is strict, their performance gap is small, indicating that data with higher entropy tends to have higher power to some extent. However, with the increase of the CBR, the gap between them begins to widen. It means that power is not a decent metric for measuring data importance compared to entropy. Data with high power is not necessarily valuable to the video communication task. In a word, data with high entropy (valuable) tends to have high power, but data with high power do not always have high entropy.

\section{Conclusion}  
The goal of this paper is to present a novel method for transmitting video semantic information over wireless channels using JSCC. We call this method MDVSC, which stands for model division video semantic communication. Unlike the conventional MEMC system combined with the channel coding approach, our method uses JSCC to avoid the cliff effect, latent space transformation to achieve better compression, model division common feature extraction to improve communication efficiency, and entropy-based variable length coding to control the communication resource cost precisely. Our method is compared with the traditional separated coding approach on different datasets and SNRs. Extended experiments and ablation studies are also conducted to understand how our method works. The results show that MDVSC outperforms the traditional wireless video communication methods in terms of PSNR and MS-SSIM. Moreover, we demonstrate that the CFE and the entropy-based variable length coding modules are beneficial for the video semantic transmission task. Our method is effective and efficient in low SNR and strict bit rate (CBR) scenarios.


%





\ifCLASSOPTIONcaptionsoff
  \newpage
\fi



\bibliographystyle{IEEEtran}
%



\bibliography{IEEEabrv, reference}

%








\end{document}